\documentclass[aps,pra,10pt,a4paper,nosuperscriptaddress,twocolumn,showpacs,titlepage,showkeys]{revtex4}
\usepackage{times}
\usepackage{amsmath}
\usepackage{amssymb}
\usepackage{epsf}
\usepackage{graphicx}
\usepackage{times}
\usepackage{amssymb}
\usepackage{graphicx}
\usepackage{amsmath}
\usepackage{epsf}
\usepackage{amsfonts}%
\begin{document}
\title{Geometric global quantum discord of two-qubit X states}
\author{Wen-Chao Qiang$^{a}$}
\thanks{Corresponding author.\\
E-mail address:qwcqj@163.com (Wen-Chao Qiang).}
\author{Hua-Ping Zhang$^a$}
\author{Lei Zhang$^b$}
\affiliation{$^a$Faculty of Science, Xi'an University of
Architecture and Technology, Xi'an, 710055, China\\
$^b$Huaqing College, Xi'an University of
Architecture and Technology, Xi'an, 710055, China}


\begin{abstract}
Xu [Jianwei Xu, J. Phys. A: Math. Theor. \textbf{45} 405304 (2012)] generalized geometric quantum discord [B. Dakic, V. Vedral, and \v{C}. Brukner, Phys. Rev. Lett. \textbf{105} 190502 (2010)] to multipartite states and proposed a geometric global quantum discord.
Almost at the same time, Hassan and Joag [J. Phys. A: Math. Theor. \textbf{45} 345301 (2012)]
introduced total quantum correlations in a general N-partite quantum state and obtained exact computable formulas for the total quantum correlations in a N-qubit quantum state.
In this paper, we pointed out that the geometric global quantum discord and the total quantum correlations are identical. We derive the analytical formulas of the geometric global quantum discord and geometric quantum discord for two-qubit $X$ states, respectively, and give five concrete examples to demonstrate the use of our formulas. Finally, we prove that the geometric quantum discord is a tight lower bound of the geometric global quantum discord.
\end{abstract}
\maketitle

\section{INTRODUCTION}\label{s1}
Quantum correlations, which are a fundamental character of a multipartite quantum system and an essential resource for quantum information processing \cite{Nielsen}, initially studied in the entanglement-versus-separability scenario \cite{pra_40_4277_1989,rmp_80_517_2008,rmp_81_865_2009}. While entanglement has attracted much effort, however, it has been found that the entanglement is
not the only characteristic of a quantum system, and it has no
advantage for some quantum information tasks. In some cases
\cite{PRA_72_042316_2005,PRL_100_050502_2008,PRL_101_200501_2008},
although there is no entanglement, certain quantum information
processing tasks can still be done efficiently by using  quantum
discord \cite{PRL_88_017901_2001,JPA_34_6899_2001,PRA_67_012320_2003},
which is believed more workable than the entanglement.
The quantum discord (QD), first introduced by Ollivier and Zurek \cite{PRL_88_017901_2001} and by Henderson and Vedral \cite{JPA_34_6899_2001}, is a measure of quantum correlations, which extends beyond entanglement, and a quantum-versus-classical paradigm for correlations
\cite{prl_100_090502_2008,pra_77_022301_2008,pra_78_024303_2008}.

In spite of its merit, because the calculation of quantum discord involves a
difficult optimization procedure, it is sometimes hard to obtain
analytical results except for a few families of two-qubit states
\cite{pra_81_042105_2010,pra_82_069902_2010,prl_89_180402_2002,prl_101_070502_2008,pra_76_032327_2007,pra_77_42303_2008,
prb_78_224413_2008,pra_80_022108_2009,prl_105_150501_2010}.
Huang have proved that computing quantum discord is NP-complete, the running time
of any algorithm for computing quantum discord is believed to grow exponentially with the
dimension of the Hilbert space. Therefore, computing quantum discord in a quantum system even with moderate size is not possible in practice \cite{njp_16_033027_2014}.
So recently, some authors restrict their research to two-qubit $X$ states, which was frequently encountered in condensed matter systems, quantum dynamics, etc. \cite{pra_81_052107_2010,prb_78_224413_2008,prl_105_150501_2010,prl_105_095702_2010,
pra_82_042316_2010}
with an interest in the dynamics of quantum discord \cite{pra_81_022116_2010}. M. Ali first studied the quantum discord for two-qubit $X$ states and derived an explicit expression for $X$ state \cite{pra_81_042105_2010,pra_82_069902_2010}, but Lu immediately gave a counterexample to Ali's results   \cite{pra_83_012327_2011}. Then Chen analyzed Ali's results and pointed out that Ali's algorithm is only valid for a class of $X$ states; there is a family of $X$ states, for which Ali's algorithm is not correct because the  inequivalence between the minimization over positive operator-valued measures and that over von Neumann measurements \cite{pra_84_042313_2011}. Soon after, Rau, one of the authors of Ref. \cite{pra_81_042105_2010, pra_82_069902_2010}, and his co-worker extend the procedure in Ref.\cite{pra_81_042105_2010} for calculating discord of two-qubit
X-states to so called extended X-states, which contain $N$ qubits. They also given a  formula to calculate the geometric measure of quantum discord for qubit-qudit systems \cite{JPA_45_095303_2012}. In this aspect, Huang also gave a counterexample to the analytical formula derived in \cite{pra_81_052107_2010}, and proposed an analytical formula with very small worst-case error \cite{pra_88_014302_2013}.

Considering the difficulty to calculate the quantum discord, Daki\'{c} \textit{et al}. \cite{prl_105_190502_2010} introduced  a geometric measure of quantum discord, which also was named geometric discord (GD), and obtained the analytic formula for two qubit states. After Daki\'{c}'s paper published soon, Luo and Fu generalized GD to an arbitrary bipartite system and derived an explicit tight lower bound on GD \cite{pra_82_034302_2010}. Rana \textit{et al}. and Hassan \textit{et al}. also independently obtained the rigorous lower bound to GD \cite{pra_85_024102_2012,pra_85_024302_2012}. D. Girolami \textit{et al}. gave another expression of GD for qubit-qubit states \cite{IJMPB_27_1345020_2012,prl_108_150403_2012}.
Tufarelli \textit{et al}. proposed an algorithm to calculate GD for any $2\times d$ systems, which is valid for $d\rightarrow\infty$ case \cite{pra_86_052326_2012}.

Because the original definitions of both QD and GD consider a set of local measurements only on one subsystem. It is not symmetrical for two subsystems in the two partite case. Rulli \textit{et al}. suggested a symmetric extension of QD named global quantum discord(GQD) \cite{PRA_84_042109_2011}, which has been extended to \textit{q}-global quantum discord \cite{PRA_87_062339_2013}. Some analytical expressions of GQD for some special quantum states also have been found \cite{pla_377_238_2013}. On the other hand, inspired by Rulli's work, Xu generalized the geometric quantum discord to multipartite states and proposed a geometric global quantum discord (GGQD)\cite{jpa_45_405304_2012}, which is alternatively called as symmetric or two-side geometric measure of quantum discord for two-qubit system\cite{PRA_86_042123_2012,cpb_22_040303_2013}. Almost meanwhile, Hassan and Joag proposed total quantum correlations (TQC) and presented an algorithm to calculate TQC for a N-partitle quantum state \cite{jpa_45_345301_2012}.
Compared with QD and GD, obviously, the study of GQD and GGQD as well as TQC is not yet enough. So, in this paper, we first prove that GGQD and TQC are identical, then we restrict ourselves to the study of GGQD. We devote to derive an  explicit analytical expression of GGQD for two-qubit $X$ states.

This paper is organized as follows. In the next section, we give a brief review of GQD and GGQD as well as TQC and prove that GGQD and TQC are identical. We derive the analytical
formulas of GGQD and GQD of two-qubit $X$ states in Sec.\ref{s3}. Some demonstrating examples are given in Sec.\ref{s4}. A related discussion is presented in Sec.\ref{s5} and we give concluding remarks in the last section.

\section{Brief review of geometric
measure of quantum discord and geometric global quantum discord}\label{s2}
For convenience of later use, we present a  brief review of QD, GD and GGQD as well as TQC.
The QD of a bipartite state $\rho$
on a system $H^a \otimes H^b$ with marginals $\rho^a$ and $\rho^b$
can be expressed as
\begin{equation}\label{qd}
   Q(\rho)=\underset{\Pi^a}{\textrm{min}}\{I(\rho)-I(\Pi^a(\rho))\}.
\end{equation}
Here the minimum is over von Neumann measurements (one-dimensional
orthogonal projectors summing up to the identity) $\Pi^a =
\{\Pi^a_k\}$ on subsystem $a$, and
\begin{equation}\label{2}
\Pi^a(\rho)=\sum_k(\Pi^a_k\otimes I^b)\rho (\Pi^a_k\otimes I^b)
\end{equation}
is the resulting state after the measurement. $I(\rho) = S(\rho^a)
+ S(\rho^b)- S(\rho)$ is the quantum mutual information, $S(\rho)
= -\textrm{tr}\rho \textrm{ln} \rho$ is the von Neumann entropy,
and $I^b$ is the identity operator on $H^b$.
The GD  for a state $\rho$ is defined as \cite{prl_105_190502_2010}:
\begin{equation}\label{Dakic}
    D(\rho)=\underset{\chi}{\textrm{min}}\|\rho-\chi\|^2,
\end{equation}
where the minimum is over the set of zero-discord states (i.e.,
$Q(\chi) = 0$) and $\|\rho-\chi\|^2:=\textrm{tr}(\rho-\chi)^2$ is
the square of Hilbert-Schmidt norm of Hermitian operators. The
GD of any two-qubit state is
evaluated as
\begin{equation}\label{D2}
    D(\rho)=\frac{1}{4}
(\|\textbf{x}\|^2+\|\textbf{T}\|^2-k_{max}),
\end{equation}
where $\textbf{x}:=(x_1,x_2,x_3)^t$ is a column vector,
$\|\textbf{x}\|^2 := \sum_i x_i^2$, $x_i=\textrm{tr}(\rho(\sigma_i
\otimes \textbf{I}^b))$, $T:=(t_{ij})$ is a matrix and
$t_{ij}=\textrm{tr}(\rho(\sigma_i \otimes \sigma_j)$, $k_{max}$ is
the largest eigenvalue of matrix $ \textbf{x}
\textbf{x}^t+\textbf{T} \textbf{T}^t$.

Since  Daki\'{c} \textit{et al}. proposed the GD,
 many authors extended Daki\'{c}'s results to the
general bipartite states. Luo and Fu evaluated the GD
 for an arbitrary state $\rho$ and obtained an
explicit formula
\begin{equation}\label{}
    D(\rho)=\textrm{tr}(\textbf{C}\textbf{C}^t)-
    \underset{A}{\text{max}}~\text{tr}(\textbf{A} \textbf{C} \textbf{C}^t
    \textbf{A}^t),
\end{equation}
where $\textbf{C} = (c_{ij})$ is an $m^2\times n^2$ matrix, given
by the expansion $\rho =\sum c_{ij} X_i\otimes Y_j$ in terms of
orthonormal operators $X_i\in L(H^a),Y_j\in L(H^b)$ and $A =
(a_{ki})$ is an $m\times m^2$ matrix given by $a_{ki} =
\textrm{tr} |k\rangle\langle k| X_i=\langle k|X_i|k\rangle$  for
any orthonormal basis ${|k\rangle}$ of $H^a$. They also gave a
tight lower bound for GD of arbitrary bipartite
states \cite{pra_82_034302_2010}. Recently, a different tight lower bound for
GD of arbitrary bipartite states was given by S.
Rana \textit{et al}. \cite{pra_85_024102_2012}, and Ali Saif M. Hassan \textit{et
al}. \cite{pra_85_024302_2012} independently. Other explicit expressions of GD
for two-qubit system and  $2\otimes d $  systems are also found  \cite{IJMPB_27_1345020_2012,prl_108_150403_2012,pra_86_052326_2012}.

Though QD and GD
 have been revealed as useful measurements, they are originally not symmetric for its all subsystems.
As an extension of QD,  Rulli
proposed a global quantum discord (GQD) for an arbitrary multipartite state
$\rho_{A_1\cdots A_N}$ as \cite{PRA_84_042109_2011}:
\begin{eqnarray}\label{QD}
  D(\rho_{A_1\cdots A_N})&=& \underset{\{\Pi_k\}}{\text{min}}[S(\rho_{A_1\cdots A_N}\| \Phi(\rho_{A_1\cdots A_N}))\nonumber\\
  &&-\sum^N_{j=1}S(\rho_{A_j}\|\Phi_j(\rho_{A_j}))].
\end{eqnarray}
To calculate  $D(\rho_{A_1\cdots A_N})$ conveniently, Xu has given an equivalent expression of Eq.(\ref{QD})\cite{pla_377_238_2013}
\begin{eqnarray}\label{QDXu}
  &&D(\rho_{A_1\cdots A_N})=\sum^N_{k=1}S(\rho_{A_k})-S(\rho_{A_1A_2\cdots A_N})\nonumber\\
  &&-\underset{\Pi}{\text{max}}\sum_{k=1}^N S(\Pi_{A_k}(\rho_{A_k}))-S(\Pi(\rho_{A_1\cdots A_N})),
\end{eqnarray}
where $\Pi=\Pi_{A_1A_2\cdots A_N}$ is any locally projective measurement performed on $A_1A_2\cdots A_n$.
Inspired by Rulli's work, Xu generalized the GD to multipartite states and proposed a geometric
 global quantum discord (GGQD) \cite{jpa_45_405304_2012},
GGQD is also called as a symmetric or two-sided geometric measure
 of quantum discord for two-qubit system \cite{PRA_86_042123_2012, cpb_22_040303_2013}.
The definition of GGQD for state $\rho_{A_1A_2 \cdots A_N}$ is
\begin{eqnarray}\label{GGQDdef}
  D^G(\rho_{A_1A_2 \dots A_N}) &=& \underset{\sigma_{A_1A_2\cdots A_N}}{\text{min}}                           \{ \text{tr}[\rho_{A_1A_2 \cdots A_N}-\sigma_{A_1A_2\cdots A_N}]^2 \nonumber\\
  && :D(\sigma_{A_1A_2\cdots A_N})=0\},
\end{eqnarray}
where $D(\sigma_{A_1A_2\cdots A_N})$ is defined by Eq.~(\ref{QD}).
To simplify calculation of Eq.(\ref{GGQDdef}), Xu derived two equivalent formulas of GGQD. The first one is:
\begin{eqnarray}\label{th1}
  &&D^G(\rho_{A_1A_2 \dots A_N}) =\underset{\Pi}{\text{min}}\{\text{tr}[\rho_{A_1A_2 \cdots A_n}-\Pi(\rho_{A_1A_2 \cdots A_N})]^2\}\nonumber\\
  &~~~~=& \text{tr}[\rho_{A_1A_2 \cdots A_N}]^2-\underset{\Pi}{\text{max}}\{ \text{tr}[\Pi(\rho_{A_1A_2 \cdots A_N})]^2\},
\end{eqnarray}
where $\Pi$ is the same as the one in Eq. (\ref{QDXu}).

The second formula of  GGQD can be expressed as
\begin{widetext}
\begin{equation}\label{th2}
D^G(\rho_{_{A_1A_2 \dots A_N}})=\sum_{\alpha_1,\alpha_2,\cdots \alpha_N}\big(C_{\alpha_1\alpha_2\cdots \alpha_N}\big)^2
-\underset{\Pi}{\text{max}}\sum_{i_1i_2\cdots i_N}\Bigg(\sum_{\alpha_1,\alpha_2,\cdots \alpha_N}A_{\alpha_1i_1}A_{\alpha_2i_2}\cdots A_{\alpha_N i_N}C_{\alpha_1\alpha_2\cdots \alpha_N}\Bigg)^2,
\end{equation}
\end{widetext}
where $C_{\alpha_1\alpha_2\cdots \alpha_N}$ and $A_{\alpha_ki_k}$ are determined by
\begin{equation}\label{C}
    \rho_{_{A_1A_2 \cdots A_N}}=\sum_{\alpha_1\alpha_2\cdots \alpha_N}C_{\alpha_1\alpha_2\cdots \alpha_N}X_{\alpha_1}\otimes X_{\alpha_2}\otimes \cdots \otimes X_{\alpha_N},
\end{equation}
\begin{equation}\label{A}
    A_{\alpha_ki_k}=\langle i_k|X_{\alpha_k}|i_k\rangle
\end{equation}
and $\{X_{\alpha_k}\}_{\alpha_k=1}^{n_k^2}$ are orthonormal bases of $L(H_k)$,
 which were constituted by all Hermitian operators on $H_k$;
 $\{|i_k\rangle\}_{i_k=1}^{n_k^2}$  are orthonormal bases of $H_k$.
 For any two-qubit state $\rho_{AB}$, Eqs.(\ref{th2}-\ref{A}) are reduced to:
 \begin{equation}\label{DGAB}
D^G(\rho_{AB})=\sum_{\alpha\beta}(C_{\alpha\beta})^2-\underset{A B}{\text{max}}\sum_{i,j}(\sum_{\alpha \beta}A_{i\alpha} B_{j\beta}C_{\alpha\beta})^2,
 \end{equation}
 \begin{equation}\label{CAB}
    C_{\alpha\beta}=\text{tr}(\rho_{AB} X_\alpha Y_\beta),
 \end{equation}
 \begin{eqnarray}\label{AB}
   A_{i\alpha}&=&\langle i|X_\alpha|i\rangle,~~B_{j\beta}=\langle j|Y_\beta|j\rangle,\nonumber \\
   &&i,j=1,2; ~~\alpha,\beta=0,1,2,3,
 \end{eqnarray}
here, for consistency with other literature, such as \cite {pra_82_034302_2010,pra_85_024302_2012}, we have exchanged the indexes
of $A$ and $B$ in Eq.(\ref{AB}),
which do not affect the latter results.
On the other hand, In Eqs.(\ref {CAB}) and (\ref{AB}), $X_0=\mathbf{I}^A/\sqrt{2},~~X_i=\sigma_i^A/\sqrt{2}, i=1,2,3;~ Y_0=\mathbf{I}^B/\sqrt{2},~~Y_j=\sigma_j^B/\sqrt{2},j=1,2,3$.
where $\mathbf{I}^A$ and $\sigma_i^A$ are $3\times 3$ unitary matrix and Pauli matrix for qubit A, $\mathbf{I}^B$ and $\sigma_j^B$ are the same
 for qubit B, respectively. We can further express Eq.(\ref{DGAB}) in the matrix form,
\begin{equation}\label{ABm}
D^G(\rho_{AB})=\text{tr}(C C^t)-\underset{A B}{\text{max}}~\text{tr}(ACB^tBC^tA^t),
 \end{equation}
where $X^t$ denote the transpose of matrix $X$, $A=\{A_{i\alpha}\}$, $B=\{B_{j\beta}\}$ and $C=\{C_{\alpha \beta}\}$.
 Equation (\ref{ABm}) is obviously the generalization of Eq.(5) in \cite{pra_82_034302_2010} to the case of GGQD.

Now, we turn our attention to TQC. Hassan and Joang introduced  total quantum correlations in a state $\rho_{12\cdots N}$  \cite{jpa_45_345301_2012}. They assumed that the non-selective von Neumann projective
measurements $\widetilde{\Pi}^{(1)}, \widetilde{\Pi}^{(2)},\cdots, \widetilde{\Pi}^{(N)}$
are acted on $N$ parts $12 \cdots N$ of the system successively. The corresponding post-measurement states are expressed by
$\widetilde{\Pi}^{(1)}(\rho_{12\cdots N}), \widetilde{\Pi}^{(2)}(\widetilde{\Pi}^{(1)}(\rho_{12\cdots N})),\cdots,\\
\widetilde{\Pi}^{(N)}(\cdots (\widetilde{\Pi}^{(1)}(\rho_{12\cdots N})\cdots ))$.
The GQD of these successive measurement states are given by
$D_1(\rho_{12\cdots N}),D_2(\widetilde{\Pi}^{(1)}(\rho_{12\cdots N})),\cdots,\\
 D_N(\widetilde{\Pi}^{(N-1)}(\cdots (\widetilde{\Pi}^{(1)}(\rho_{12\cdots N}))\cdots ))$.
Then, the geometric measure of total quantum correlations of a N-partite quantum state
$\rho_{12\cdots N}$ is defined as
\begin{eqnarray}\label{Q}
  Q( \rho_{12\cdots N})&=& D_1(\rho_{12\cdots N})+D_2(\widetilde{\Pi}^{(1)}(\rho_{12\cdots N}))+\cdots \nonumber\\
   &+& D_N(\widetilde{\Pi}^{(N-1)}(\cdots (\widetilde{\Pi}^{(1)}(\rho_{12\cdots N}))\cdots )).
\end{eqnarray}
In the following, we shall see that the definitions of GGQD and TQC are different in form,
but they are identical to each other. For this end, we recall that, obviously, Eq.(\ref{th1}) is also valid for GQD with $\Pi=\Pi^{k},k=1,2,\cdots,N$, which only performed on $k$th part of the system. Keeping this in mind, we can rewrite Eq.(\ref{Q}) for $N=2$ as
\begin{eqnarray}\label{Q=DG}
  Q(\rho) &=& D_1(\rho)+D_2(\widetilde{\Pi}^{(1)}(\rho)) \nonumber\\
   &=& \text{tr}[\rho]^2-\underset{\Pi^{(1)}}{\text{max}}\{ \text{tr}[\Pi^{(1)}(\rho)]^2\}\nonumber\\
   & &+
   \text{tr}[\Pi^{(1)}(\rho)]^2-\underset{\Pi^{(2)}}{\text{max}}\{ \text{tr}[\Pi^{(2)}(\Pi^{(1)}(\rho))]^2\}\nonumber\\
   &=&\text{tr}[\rho]^2-\underset{\Pi^{(2)}\Pi^{(1)}}{\text{max}}\{\text{tr}[\Pi^{(2)}(\Pi^{(1)}(\rho))]^2\}\nonumber\\
   &=&\text{tr}[\rho]^2-\underset{\Pi}{\text{max}}\{\text{tr}[\Pi(\rho)]^2\}=D^G(\rho)
\end{eqnarray}
In the above equation, $\Pi=\Pi^{(2)}\Pi^{(1)}$.  The terms $\underset{\Pi^{(1)}}{\text{max}}\{ \text{tr}[\Pi^{(1)}(\rho)]^2\}$ and
   $\text{tr}[\Pi^{(1)}(\rho)]^2$ after the second equal sign are canceled because $\Pi^{(1)}$ maximize $\text{tr}[\Pi^{(1)}(\rho)]^2$. The proof of $Q(\rho)=D^G(\rho)$ for $N\geq 3$ cases is similar and straightforward. The identity of GGQD with TQC is not surprising, because these two measurements both use the original definition of the geometric measure of quantum discord to every individuals of the system.  Due to this identity,                                                                                                                                                                                                                                                                                                                        therefore, hereafter we use  the name 'geometric global quantum discord (GGQD)', which also stand for 'total quantum correlations (TQC)'. In the next section, we use Eq.(\ref{ABm}) to calculate the GGQD of $X$ state.

\section{GGQD OF TWO-QUBIT $X$ STATE}\label{s3}
The two-qubit X state usually arises as the two-particle
reduced density matrix in many physical systems. In the computational basis ${|00\rangle,|01\rangle,|10\rangle,|11\rangle}$
, the visual appearance of its density matrix  resembles the letter X leading it  to be called as X state. The density matrix of a two-qubit X state
\begin{equation}\label{rox}
    \rho_{AB}=\left(
       \begin{array}{cccc}
         \varrho_{00} & 0 & 0 & \varrho_{03} \\
         0 & \varrho_{11} & \varrho_{12} & 0 \\
         0 & \varrho_{12}^* & \varrho_{22} & 0 \\
         \varrho_{03}^* & 0 & 0 &  \varrho_{33}\\
       \end{array}
     \right)
\end{equation}
has nonzero elements only on the diagonal and the antidiagonal,
where $\varrho_{00},\varrho_{11},\varrho_{22},\varrho_{33}\geq 0$ satisfy $\varrho_{00} +\varrho_{11} +\varrho_{22} +\varrho_{33} = 1 $. The antidiagonal elements $\varrho_{03},\varrho_{12}$ are generally complex
numbers, but can be made real and nonnegative by the
local unitary transformation $ e^{-i \theta_1 \sigma_z}\otimes  e^{-i \theta_2 \sigma_z}$ with $\theta_1=-(\arg{\varrho_{03}}+\arg{\varrho_{12}})/4,\theta_2=-(\arg{\varrho_{03}}-\arg{\varrho_{12}})/4$,
where $\sigma$ is the Pauli matrix. Hereafter we assume $\varrho_{03},\varrho_{12}\geq 0$.
Recall that the matrix $C$ in Eq.(\ref{ABm}) can be written as \cite{pra_82_034302_2010,pra_85_024102_2012}
\begin{equation}\label{cc}
  C=(C_{ij})=\frac{1}{2}
\left(
  \begin{array}{cc}
    1 & \textbf{y}^t \\
    \textbf{x} & T \\
  \end{array}
\right),
\end{equation}
Matrix $A$ and $B$ in Eq.(\ref{ABm}) can be expressed as \cite{pra_82_034302_2010}
\begin{equation}\label{AAA}
    A=\frac{1}{\sqrt{2}}
\left(
  \begin{array}{cc}
    1 & \textbf{a} \\
    1 & -\textbf{a} \\
  \end{array}
\right),
\end{equation}
\begin{equation}\label{BBB}
    B=\frac{1}{\sqrt{2}}
\left(
  \begin{array}{cc}
    1 & \textbf{b} \\
    1 & -\textbf{b} \\
  \end{array}
\right),
\end{equation}
where $\textbf{a}=\{a_1,a_2,a_3\}=\sqrt{2}(A_{11},A_{12},A_{13})$, $\textbf{b}=\{b_1,b_2,b_3\}=\sqrt{2}(B_{11},B_{12},B_{13})$ and
$\|\textbf{a}\|=\|\textbf{b}\|=1$.
Using Eqs.(\ref{cc}-\ref{BBB}), we can easily get the first term in Eq.(\ref{ABm})
\begin{equation}\label{trcc}
    \text{tr}(CC^t) = \frac{1}{4}(1+\|\textbf{x}\|^2+\|\textbf{y}\|^2+\|T\|^2)
\end{equation}
and the second term
\begin{equation}\label{TrACB}
   \text{tr}(ACB^tBC^tA^t) =\frac{1}{4}\big[1
+\textbf{y}^t\textbf{b}^t\textbf{b}\textbf{y}+\textbf{a}(\textbf{x}\textbf{x}^t+T\textbf{b}^t \textbf{b}T^t)\textbf{a}^t\big].
\end{equation}
The maximum over matrixes $A$ and $B$ in Eq.(\ref{ABm}) can be done by two steps.
First, we maximize $\textbf{a}(\textbf{x}\textbf{x}^t+T\textbf{b}^t \textbf{b}T)\textbf{a}^t$ on matrix $A$.
The maximum of this term is the largest eigenvalue $\lambda_{max-A}$ of matrix $\textbf{x}\textbf{x}^t+T\textbf{b}^t \textbf{b}T$. According to the Lemma 1 of
Ref.\cite{cpb_22_040303_2013}, which states that for any two vectors $|a\rangle$ and  $|b \rangle$  (not necessarily
normalized) in $\mathbb{R}^3$, the largest eigenvalue of the matrix $|a\rangle \langle a|+|b\rangle \langle b|$ is
$\lambda=[a^2+b^2+\sqrt{(a^2-b^2)^2+4 \langle a|b\rangle^2}]/2$
with $a^2 =\langle a|a\rangle$ and $b^2 =\langle b|b\rangle$, we get
\begin{equation}\label{lamda_A}
   \lambda_{max-A}= \frac{1}{2}[\|\textbf{x}\|^2+\|\textbf{b} T^t\|^2+\sqrt{(\|\textbf{x}\|^2-\|\textbf{b} T^t\|^2)^2+4 (\textbf{x}^t \textbf{b} T^t)^2}].
\end{equation}
Substituting Eqs.(\ref{trcc} - \ref{lamda_A}) into Eq.(\ref{ABm}), we obtain the GGQD of any two-qubit systems
\begin{widetext}
\begin{equation}\label{GGQB}
   D^G(\rho_{AB})= \frac{1}{4}\big\{\|\textbf{x}\|^2+\|\textbf{y}\|^2+\|T\|^2 -\frac{1}{2} \underset{ \textbf{b}}{\text{max}}[\|\textbf{x}\|^2+\|\textbf{b} T^t\|^2+\sqrt{(\|\textbf{x}\|^2-\|\textbf{b} T^t\|^2)^2+4 (\textbf{x}^t \textbf{b} T^t)^2}+2 \|\textbf{by}\|^2]\big\}.
\end{equation}
\end{widetext}
The second step to maximize $\text{tr}(ACB^tBC^tA^t)$ in Eq.(\ref{ABm}) is reduced to maximize $\|\textbf{x}\|^2+\|\textbf{b} T^t\|^2+\sqrt{(\|\textbf{x}\|^2-\|\textbf{b} T^t\|^2)^2+4 (\textbf{x}^t \textbf{b} T^t)^2}+2 \|\textbf{by}\|^2$ in above equation on $\textbf{b}=\{b_1,b_2,b_3\}$ .
For $X$ state (\ref{rox}),
\begin{eqnarray}
  \textbf{x}^t &=& \{x_1,x_2, x_3\}=\{0, 0, \varrho_{00} +\varrho_{11} -\varrho_{22} -\varrho_{33}\}, \\
  \textbf{y}^t & =& \{y_1,y_2,y_3\} =\{0, 0, \varrho_{00} -\varrho_{11} +\varrho_{22} -\varrho_{33}\},
\end{eqnarray}
\begin{eqnarray}
  T &=& \left(
        \begin{array}{ccc}
          T_{11} & 0 & 0 \\
          0 & T_{22}& 0 \\
          0 & 0 & T_{33} \\
        \end{array}
      \right) \nonumber\\
   &=& \left(
        \begin{array}{ccc}
          2( \varrho_{12}+\varrho_{03}) & 0 & 0 \\
          0 & 2( \varrho_{12}-\varrho_{03})& 0 \\
          0 & 0 & \varrho_{00} -\varrho_{11} -\varrho_{22}+\varrho_{33} \\
        \end{array}
      \right),\nonumber\\
\end{eqnarray}
\begin{equation}\label{ABC}
\begin{split}
&\|\textbf{x}\|^2 + \|\textbf{b} T^t\|^2+2\|\textbf{by}\|^2+\sqrt{(\|\textbf{x}\|^2-\|\textbf{b} T^t\|^2)^2+4 (\textbf{x}^t \textbf{b} T^t)^2}  \\
  &=x_3^2+V+ 2 b_3^2y_3^2
+\sqrt{\left(x_3^2-V\right)^2-4 W x_3^2},
\end{split}
\end{equation}
where
\begin{eqnarray}
W &=& b_1^2 T_{11}^2 + b_2^2 T_{22}^2,~~~V= b_3^2 T_{33}^2+ W.
\end{eqnarray}
To further maximize Eq.(\ref{ABC}) we let $$b_1=\sin\theta \cos\phi,~b_2=\sin\theta \sin\phi,
~b_3= \cos \theta,$$ Eq.(\ref{ABC}) becomes
\begin{eqnarray}
  f(\theta,\phi) &=& \frac{1}{2} \left[x_3^2+2 y_3^2 \cos^2\theta
   +\gamma (\theta,\phi)\right. \nonumber \\
  & & \left. +\sqrt{(\gamma (\theta ,\phi)
   -x_3^2)^2+4 T_{33}^2 x_3^2 \cos
   ^2\theta)}\right],  \nonumber \\
  \gamma(\theta,\phi)&=&
    T_{33}^2\cos^2\theta+\mu(\phi)\sin^2\theta, \\
\mu(\phi)&=&T_{11}^2\cos^2\phi +
     T_{22}^2\sin^2\phi. \nonumber
\end{eqnarray}
Ignoring the relative maximum $x_3^2$ of $f(\theta,\phi)$, we find
\begin{equation}\label{df}
    \begin{array}{ll}
  \{\frac{\partial f(\theta,\phi)}{\partial \theta},\frac{\partial f(\theta,\phi)}{\partial \phi}\}|_{\theta=0}=0, & f(0,\phi)= x_3^2 + y_3^2+T_{33}^2,\\
   \{\frac{\partial f(\theta,\phi)}{\partial \theta},\frac{\partial f(\theta,\phi)}{\partial \phi}\}|_{\theta=\frac{\pi}{2},\phi=0\bigvee\pi}=0, & f(\frac{\pi}{2},0\bigvee \pi)=T_{11}^2, \\
\{\frac{\partial f(\theta,\phi)}{\partial \theta},\frac{\partial f(\theta,\phi)}{\partial \phi}\}|_{\theta=\frac{\pi}{2},\phi=\frac{\pi}{2}\bigvee \frac{3\pi}{2}}=0, &
f(\frac{\pi}{2},\frac{\pi}{2}\bigvee \frac{3\pi}{2})=T_{22}^2, \\
\end{array}.
\end{equation}
Finally, we obtain the maximum value of Eq.(\ref{ABC})
$$\text{max}[\varrho_{00}^2 + \varrho_{11}^2 + \varrho_{22}^2 + \varrho_{33}^2 - 1/4, (\varrho_{12} +\varrho_{03})^2]$$
and the GGQD of $X$ states
\begin{eqnarray}\label{GDX}
 D^G(\rho_{X})&=&\varrho_{00}^2 + \varrho_{11}^2 + \varrho_{22}^2 + \varrho_{33}^2 - \frac{1}{4}+2(\varrho_{12}^2+\varrho_{03}^2) \nonumber\\
&-&\text{max}[\varrho_{00}^2 + \varrho_{11}^2 + \varrho_{22}^2 + \varrho_{33}^2 - \frac{1}{4}, (\varrho_{12} +\varrho_{03})^2].\nonumber\\
\end{eqnarray}
For comparing GGQD with GD for some $X$ states in next section, we also calculated the GD of $X$ state according to
    Ref.\cite{pra_82_034302_2010}
and got the following formula:
\begin{eqnarray}\label{DX}
 D(\rho_{X})&=& \frac{1}{2}(\varrho_{00}^2 + \varrho_{11}^2 + \varrho_{22}^2 + \varrho_{33}^2) \nonumber\\
 &-& \varrho_{00} \varrho_{22} - \varrho_{11}\varrho_{33}
+2(\varrho_{12}^2+\varrho_{03}^2) \nonumber\\
&-&\text{max}[\frac{1}{2}(\varrho_{00}^2 + \varrho_{11}^2 + \varrho_{22}^2 + \varrho_{33}^2) \nonumber\\
&-& \varrho_{00} \varrho_{22} - \varrho_{11}\varrho_{33},
 (\varrho_{12} +\varrho_{03})^2].
\end{eqnarray}
In simplifying Eqs.(\ref{GDX},\ref{DX}) the condition $\varrho_{00} + \varrho_{11} + \varrho_{22} + \varrho_{33}=1$ was repeatedly  used. This formula can also be derived by Eq.(23) of Ref.\cite{JPA_45_095303_2012}.

\section{ILLUSTRATIVE EXAMPLES}\label{s4}
In this section, we give some concrete examples to demonstrate the use of formulas obtained in above section.

\begin{figure}[htb]
\centering
\includegraphics[width=6.5cm]{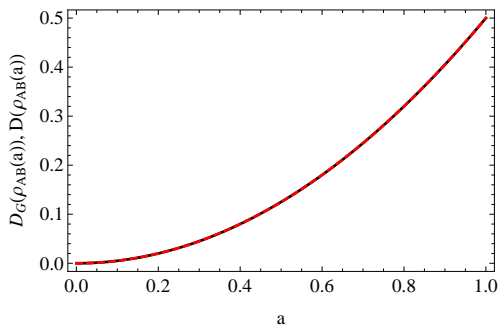}
\caption{(Color online) Graphs of  $D^G(\rho_X)$(black line) and $D(\rho_X)$(dashed and red line) as functions of the parameter $a$ for the class of states in Eq. (\ref{e1}).} \label{F1}
\end{figure}

(1) As the first example, we consider the initial state $\rho = a|\phi^+\rangle \langle \phi^+|+(1-a)|1_A,1_B\rangle \langle 1_A,1_B\rangle (0<a\leq 1)$, where $|\phi^+\rangle=(|0_A,0_B\rangle+|1_A,1_B\rangle)/\sqrt{2}$ is a maximally entangled state
\cite{pra_81_042105_2010}. The density matrix of this state is:
\begin{equation}\label{e1}
   \rho_X=\left(
          \begin{array}{cccc}
          \frac{a}{2} & 0 & 0 & \frac{a}{2} \\
         0 & 0 & 0 & 0\\
0 & 0 & 0 & 0\\
 \frac{a}{2} & 0 & 0 &  1 - \frac{a}{2} \\
          \end{array}
        \right).
\end{equation}
The corresponding GGQD and GD are
\begin{equation}\label{GGD-GD}
    D^G(\rho_X)=D(\rho_X)=\frac{a^2}{2}.
\end{equation}
We plot $D^G(\rho_X)$ and $D(\rho_X)$ in Fig.\ref{F1}. We noticed that GGQD and GD are completely  coincident in this state.

\begin{figure}[htb]
\centering
\includegraphics[width=6.5cm]{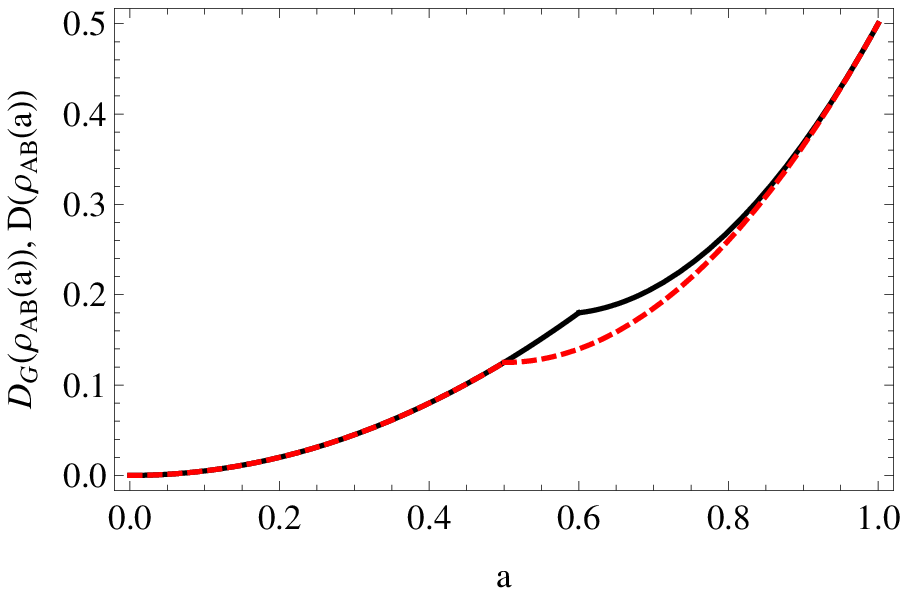}
\caption{(Color online) Graphs of  $D^G(\rho_X)$(black line) and $D(\rho_X)$(dashed and red line) as functions of the parameter $a$ for the class of states in Eq.(\ref{GD2}).} \label{F2}
\end{figure}

(2) We take the class of states defined as $\rho=a|\psi^+\rangle \langle \psi^+|+(1-a)|1_A,1_B\rangle \langle 1_A,1_B\rangle (0\leq a\leq 1)$,where
$|\psi^+\rangle=(|0_A,1_B\rangle+|1_A,0_B\rangle)/\sqrt{2}$ is a maximally entangled state
\cite{pra_81_042105_2010}. The density matrix of this state is:
\begin{equation}\label{GD2}
   \rho_X=\left(
 \begin{array}{cccc}
          0  & 0 & 0 & 0\\
0 & \frac{a}{2} & \frac{a}{2} & 0\\
0 & \frac{a}{2} & \frac{a}{2} & 0\\
0 & 0 & 0 & 1 - a\\
 \end{array}
        \right).
\end{equation}
The corresponding GGQD and GD are
\begin{equation}\label{GD2_1}
    D^G(\rho_X) = \left\{\begin{array}{ll}
\frac{a^2}{2}, & 0 \leq a \leq \frac{3}{5}   \\
 \frac{1}{4} (3 - 8 a + 7 a^2), & \frac{3}{5} < a \leq 1.\\
\end{array}\right.
\end{equation}
\begin{equation}\label{GD2_2}
    D(\rho_X)=\left\{\begin{array}{ll}
\frac{a^2}{2}, & 0 \leq a \leq \frac{1}{2}   \\
 \frac{1}{2} (1 - 3 a + 3 a^2), & \frac{1}{2} < a \leq 1.\\
\end{array}\right.
\end{equation}
We plot $D^G(\rho_X)$ and $D(\rho_X)$ for the state (\ref{GD2}) in Fig.\ref{F2}. We see that $D^G(\rho_X)=D(\rho_X)$, for $0\leq a \leq \frac{1}{2}$ and $D^G(\rho_X)\geq D(\rho_X)$, for $ \frac{1}{2} < a \leq 1$. Finally $D^G(\rho_X)= D(\rho_X)$ when $a=1$.

\begin{figure}[htb]
\centering
\includegraphics[width=6.5cm]{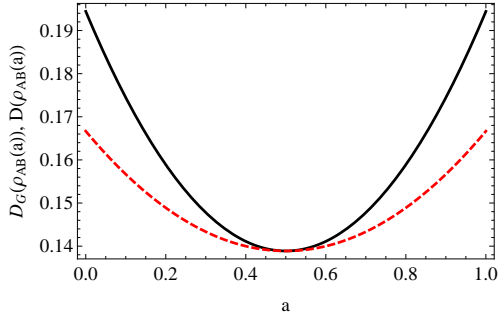}
\caption{(Color online)(Color online) Graphs of  $D^G(\rho_X)$(black line) and $D(\rho_X)$(dashed and red line)  as functions of the parameter $a$ for the class of states in Eq.(\ref{e5}).} \label{F3}
\end{figure}

(3) We take the class of states defined as $\rho=\frac{1}{3}\{(1-a)|0_A,0_B\rangle \langle 0_A,0_B|+2 |\psi^+\rangle \langle \psi^+|+a|1_A,1_B\rangle \langle 1_A,1_B\rangle (0\leq a\leq 1)$,where
$|\psi^+\rangle $ is the same as in example (2)
\cite{pra_81_042105_2010}. The density matrix of this state is:
\begin{equation}\label{e5}
   \rho_X =\left(
 \begin{array}{cccc}
          \frac{1-a}{3}  & 0 & 0 & 0\\
0 & \frac{1}{3} & \frac{1}{3} & 0\\
0 & \frac{1}{3} & \frac{1}{3} & 0\\
0 & 0 & 0 & \frac{a}{3}\\
 \end{array}
        \right).
\end{equation}
The corresponding GGQD and GD are
\begin{equation}\label{GD2_1}
    D^G(\rho_X) = \frac{1}{36}(7 - 8 a + 8 a^2),
\end{equation}
\begin{equation}\label{GD2_2}
    D(\rho_X)=\frac{1}{18} (3 - 2 a + 2 a^2).
\end{equation}
We plot $D^G(\rho_X)$ and $D(\rho_X)$ for the state (\ref{e5}) in Fig.\ref{F3}. We see that $D^G(\rho_X)$ and $D(\rho_X)$ have the same minimum values $\frac{5}{36}$ at $a=\frac{1}{2}$. The two curves are symmetrical about $a=\frac{1}{2}$.

\begin{figure}[htb]
\centering
\includegraphics[width=6.5cm]{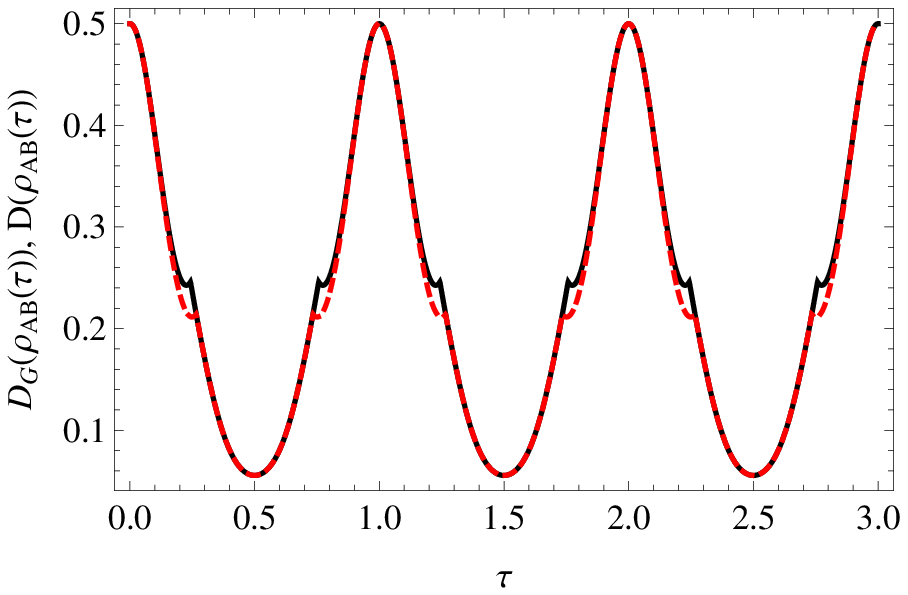}
\caption{(Color online) The evolution of $D^G(\rho_{AB})$ and $D(\rho_{AB})$ as functions of the dimensionless time $\tau=\sqrt{6}gt/(2 \pi)$ for the initial state
$|\psi(0)\rangle = (\alpha|0_{_A} 0_ {_B}\rangle + \beta|1_{_A}
1_{_B}\rangle)|n_{_C}\rangle$ with $\alpha =\beta = 1/\sqrt{2}$.
The red solid line corresponds to $D^G(\rho_{AB})$ and the blue dashed line to
$D(\rho_{AB})$.} \label{F4}
\end{figure}

(4) Two atoms in the Tavis-Cumming model \cite{pra_85_022320_2012}.
We consider two atoms (A and B), each
of which interacting resonantly with a single quantized
cavity field (system C) in a Fock state. This system is described by the two-atom Tavis-Cummings (TC) Hamiltonian:
 $H
=\hbar g[(\sigma_A + \sigma_B)a_C^\dag + (\sigma_A^\dag +
\sigma_B^\dag)a_C]$, where $\sigma_j$ and $\sigma_j^\dag$ denote the
Pauli ladder operators for the $j$th atom, $a (a^\dag)$ stands for the
annihilation (creation) operator of photons in cavity $C$, and
$g$ is the coupling constant. We consider that the system is initially in
the state $|\psi(0)\rangle = (\alpha|0_{_A} 0_ {_B}\rangle +
\beta|1_{_A} 1_{_B}\rangle)|n_{_C}\rangle$. Because the total number of excitations is conserved by TC Hamiltonian, the cavity mode will develop within a five-dimensional Hilbert space spanned
by $\{|(n-2)_{_C}\rangle,|(n-1)_{_C}\rangle,|n_{_C}\rangle,|(n +
1)_{_C}\rangle,|(n + 2)_{_C}\rangle\}$ for $n\geq 2$. When $n =
0,1$ the dimension will be $3$ and $4$, respectively.
 On the other hand, since the atomic system evolves within the subspace
$\{|0_{_A}0_{_B}\rangle,|+\rangle,|1_{_A}1_{_B}\rangle\}$ with
$|+\rangle = (|1_{_A}0_{_B}\rangle +
|0_{_A}1_{_B}\rangle)/\sqrt{2}$ independently of $n$, for our purpose,  we only need to consider $n=0$ case.
Solving the Schr\"{o}dinger equation, we obtain the state of the system at time $t$,
\begin{eqnarray}
  |\psi(t)\rangle &=& c_1(t)|0_{_A} 0_{_B}\rangle|2_{_C}\rangle + c_2(t )|+\rangle|
1_{_C}\rangle   \nonumber \\
   && +c_3(t )|1_{_A}1_{_B}|0_{_C}\rangle +
c_4(t)|0_{_A}0_{_B}\rangle|0_{_C}\rangle
\end{eqnarray}
with the following probability amplitudes
\begin{eqnarray}
  c_1(t)&=&-\frac{\sqrt{2}}{3} \beta [1 - \cos(\sqrt{6} g t)], \nonumber\\
  c_2(t)&=&-\frac{i \beta}{\sqrt{3}}\sin(\sqrt{6} g t),  \nonumber\\
  c_3(t)&=&\beta\left\{1 + \frac{1}{3} [\cos(\sqrt{6} g t)-1]\right\},\nonumber\\
  c_4(t)&=& \alpha.
\end{eqnarray}
Now, we take trace of density operator
$\rho=|\psi(t)\rangle\langle \psi(t)|$ over cavity $C$ resulting in
the reduced density matrix of the qubit-qubit system
\begin{equation}\label{TCrAB}
    \rho_{AB}=\left(
                \begin{array}{cccc}
                  |c_1|^2 +|c_4|^2 & 0 & 0 & |c_3 c_4|\\
                  0 & \frac{|c_2|^2}{2} & \frac{|c_2|^2}{2} & 0\\
                  0 & \frac{|c_2|^2}{2} & \frac{|c_2|^2}{2} & 0\\
                 |c_3 c_4| & 0 & 0 & |c_3|^2\\
                \end{array}
              \right).
\end{equation}
Using Eqs.(\ref{GDX}) and (\ref{DX}), we obtain
\begin{widetext}
\begin{equation}\label{GD_C1}
    D^G(\rho_{AB})=(|c_1|^2 +|c_4|^2)^2 + |c_2|^4 + |c_3|^4 +
 2 |c_3c_4|^2 -\frac{1}{4} -\text{max}[\frac{1}{2}|c_2|^4 + |c_3|^4 + (|c_1|^2 + |c_4|^2)^2
-\frac{1}{4}, (\frac{1}{2}|c_2|^2 + |c_3 c_4|)^2],
\end{equation}
\begin{eqnarray}\label{D_C1}
  D(\rho_{AB}) &=& \frac{1}{2} (|c_1|^4 + 4 |c_2|^4 + |c_3|^4 + |c_4|^4 -
     |c_2|^2) + (|c_1|^2 + 2 |c_3|^2) |c_4|^2 \nonumber\\
  &-&  \text{max}[\frac{1}{2} (\frac{1}{2}|c_2|^4 - 1 - |c_2|^2) +
       (1 - |c_2|^2)|c_3|^2 , (\frac{1}{2}|c_2|^2 + |c_3 c_4|)^2].
\end{eqnarray}
\end{widetext}
In this case, $D^G(\rho_{AB})$ and $D(\rho_{AB})$ as functions of dimensionless time $\tau=\sqrt{6}gt/(6\pi)$ are plotted in Fig.\ref{F4}, which shows that $D^G(\rho_{AB})$ and $D(\rho_{AB})$ change periodically with a period $T_\tau=1$. In addition, they simultaneously arrive their maximums and minimums. Furthermore, the practical calculation shows the results for $n\geq 1$ are the same as Fig.\ref{F4}, which enhances that the evolution of two atomic system is independent of $n$, as pointed out earlier.

\begin{figure}[htb]
\centering
\includegraphics[width=6.5cm]{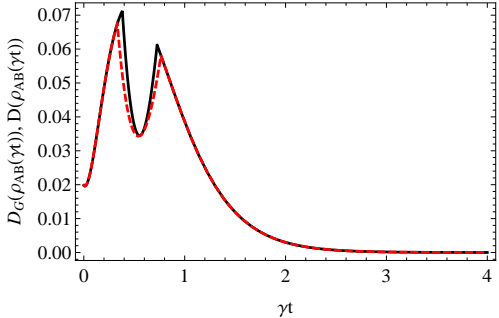}
\caption{(Color online)  The evolution of $D^G(\rho_{AB})$ and $D(\rho_{AB})$ as functions of the dimensionless time $\gamma t$ for the initial state
$|\psi(0)\rangle = (\alpha|0_{_A} 0_ {_B}\rangle + \beta|1_{_A}
1_{_B}\rangle)|\bar{0}_{_C}\rangle$ with $\alpha =0.1$ and $\beta=\sqrt{1-\alpha^2}$.
The red solid line corresponds to $D^G(\rho_{AB})$ and the blue dashed line to
$D(\rho_{AB})$.} \label{F5}
\end{figure}

(5) As a final example, let us consider  two
atoms $A$ and $B$ in  a common reservoir $C$ \cite{pra_85_022320_2012}. We suppose that the initial state of this system was $|\Psi(0)\rangle =(¦Á|g_Ag_B\rangle + ¦Â|e_Ae_B\rangle)|\bar{0}\rangle_C,$ where $|\bar{0}\rangle = \prod_k |0\rangle_k$ is the reservoir vacuum state. The overall state of the system at time $t$ can be written as
\begin{eqnarray}\label{TCrAB2}
  |\Psi(t)\rangle &=& \alpha |g_Ag_B\rangle |\bar{0}_C\rangle+c_1(t)|e_Ae_B\rangle |\bar{0}_C\rangle \nonumber \\
   &=& c_2(t)|+\rangle_{AB}|\bar{1}_C\rangle+c_3(t)|g_Ag_B\rangle |\bar{2}_C\rangle,
\end{eqnarray}
where $|+\rangle_{AB}=(|e_Ag_B\rangle+|g_Ae_B\rangle)/\sqrt{2}$ and $|\bar{k}\rangle$ denotes the collective states of the reservoir in $k$ excitations. The probability amplitudes for  this case are
\begin{eqnarray}
  c_1(t) &=& \beta e^{-\gamma t},~~~c_2(t)=\beta \sqrt{2 \gamma t} e^{-\gamma t},\nonumber \\
  c_3(t) &=& \sqrt{1-\alpha^2-c_1^2(t)-c_2^2(t)}.
\end{eqnarray}
Tracing out the reservoir, we obtain the density matrix of atomic subsystem
\begin{equation}\label{TCr2ABm}
  \rho_{AB}=\left(
                \begin{array}{llll}
                  \alpha^2+c_3^2  & 0 & 0 & \alpha c_1 \\
                  0 & \frac{c_2^2}{2} & \frac{c_2^2}{2} & 0\\
                  0 & \frac{c_2^2}{2} & \frac{c_2^2}{2} & 0\\
                 \alpha c_1  & 0 & 0 & c_1^2
                \end{array}
              \right),
\end{equation}
which is just an $X$ state. The corresponding GGQD and GD are
\begin{eqnarray}
  D^G(\rho_{AB}) &=& 3/4 - 2 [c_1^2 (c_2^2 + c_3^2) +
     c_2^2 (c_3^2 + \alpha^2)]\nonumber\\
   &-&\text{max}[ c_1^4 + c_2^4/2
    + (\alpha^2+c_3^2)^2-1/4,\\
  && (c_2^2/2 + \alpha c_1)^2], \nonumber
\end{eqnarray}
\begin{eqnarray}
  D(\rho_{AB}) &=& \frac{1}{4}[2 + 7 c_2^4 - 6 c_2^2 -
      4 c_1^2 (c_3^2 - \alpha^2)]\nonumber\\
  &-& \text{max}\{\frac{1}{2} [1 - 3 c_2^2 (1 - c_2^2)- 2 c_1^2 (c_3^2 + \alpha^2) - c_2^4/
       2],\nonumber\\
  &&(c_2]^2/2 +c_1 \alpha)^2\}.
\end{eqnarray}
In deducing above two equations, $c_1^2+c_2^2+c_3^2+\alpha^2=1$ has been used. We plot $D^G(\rho_{AB})$
and $D(\rho_{AB})$ as functions of the dimensionless time $\gamma t$ in Fig.\ref{F5}.
$D^G(\rho_{AB})$ and $D(\rho_{AB})$
have the same initial values $2\alpha^2(1 - \alpha^2)$
and two relative maximums as well as one relative minimum, respectively.
Their  corresponding relative maximums and relative minimums  are close to each other. When $t\rightarrow\infty$, $D^G(\rho_{AB})$ and $D(\rho_{AB})$ simultaneously go to zero.

\section{DISCUSSION}\label{s5}
We have derived analytical formulas of GGQD and GD for two-qubit $X$ states. Here we give some useful remarks. First, it should be pointed out that Eqs.(\ref{ABm},\ref{GGQB}), from which Eq.(\ref{GDX}) was derived, not only applicable to two-qubit $X$ states, but also to any two-qubit states. Second, because of $ \text{tr}(ACB^tBC^tA^t) = \text{tr}(BC^tA^tACB^t)$, we can alternatively first optimize system $B$, then system $A$. This is equivalent to exchange subsystems $A$ and $B$, and transpose matrix $C$. Of course, the two procedures give the same results.
Third, and more important, we find that GGQD are always greater than or equal to GD in five examples we have given. In fact, this is true for any $X$ state. We present a proof as  follows.

First, using $\text{tr}(\rho_X)=\varrho_{00}+\varrho_{11}+\varrho_{22}+\varrho_{33}=1$ we easily obtain
\begin{eqnarray}\label{DG_D}
&&(\varrho_{00}^2+\varrho_{11}^2+\varrho_{22}^2+\varrho_{33}^2 - 1/4)\nonumber\\
&-&[(\varrho_{00}^2+\varrho_{11}^2+\varrho_{22}^2+\varrho_{33}^2)/2- \varrho_{00} \varrho_{22} - \varrho_{11} \varrho_{33}]  \nonumber\\
&=&[2 (\varrho_{00} +  \varrho_{22})-1]^2/4 =[2 (\varrho_{11} + \varrho_{33})-1]^2/4\geq 0,\nonumber\\
\end{eqnarray}
which means $\varrho_{00}^2+\varrho_{11}^2+\varrho_{22}^2+\varrho_{33}^2 - 1/4\geq(\varrho_{00}^2+\varrho_{11}^2+\varrho_{22}^2+\varrho_{33}^2)/2- \varrho_{00} \varrho_{22} - \varrho_{11} \varrho_{33}$. Therefore, there are only three cases need to be considered.\\
(1)~$(\varrho_{12}+\varrho_{03})^2\geq \varrho_{00}^2+\varrho_{11}^2+\varrho_{22}^2+\varrho_{33}^2 - 1/4\geq(\varrho_{00}^2+\varrho_{11}^2+\varrho_{22}^2+\varrho_{33}^2)/2- \varrho_{00} \varrho_{22} - \varrho_{11} \varrho_{33} $:
\begin{eqnarray}
D^G(\rho_{AB})&=& \varrho_{00}^2+\varrho_{11}^2+\varrho_{22}^2+\varrho_{33}^2 -\frac{1}{4}\nonumber \\
&+& (\varrho_{12}-\varrho_{03})^2, \\
D(\rho_{AB})&=& \frac{1}{2} [(\varrho_{00} - \varrho_{22})^2 + (\varrho_{11} - \varrho_{33})^2]\nonumber \\
&+&(\varrho_{12}-\varrho_{03})^2,
\end{eqnarray}
\begin{eqnarray}\label{DG_D1}
                D^G(\rho_{AB})-D(\rho_{AB})&=&\frac{1}{4} [ 2(\varrho_{00} + \varrho_{22})-1]^2\nonumber \\
               &=&\frac{1}{4} [ 2(\varrho_{11} + \varrho_{33})-1]^2\geq 0.
\end{eqnarray}
(2)~$\varrho_{00}^2+\varrho_{11}^2+\varrho_{22}^2+\varrho_{33}^2 - 1/4 \geq (\varrho_{12}+\varrho_{03})^2\geq(\varrho_{00}^2+\varrho_{11}^2+\varrho_{22}^2+\varrho_{33}^2)/2- \varrho_{00} \varrho_{22} - \varrho_{11} \varrho_{33} $:
\begin{eqnarray}
  D^G(\rho_{AB}) &=& 2 (\varrho_{12}^2+\varrho_{03}^2),\\
  D(\rho_{AB}) &=&\frac{1}{2} [(\varrho_{00} -\varrho_{22})^2 + (\varrho_{11}-\varrho_{33})^2]\nonumber\\
& &+ (\varrho_{12}-\varrho_{03})^2,
\end{eqnarray}
\begin{eqnarray}\label{DG_D2}
  &&D^G(\rho_{AB})-D(\rho_{AB}) \nonumber\\
   &=&(\varrho_{12}+\varrho_{03})^2
-\frac{1}{2} [(\varrho_{00} -\varrho_{22})^2 + (\varrho_{11}-\varrho_{33})^2]\nonumber\\
 &\geq &  \frac{1}{2}(\varrho_{00}^2+\varrho_{11}^2+\varrho_{22}^2+\varrho_{33}^2)
 - \varrho_{00} \varrho_{22} - \varrho_{11} \varrho_{33}\nonumber\\
&-&\frac{1}{2} [(\varrho_{00} -\varrho_{22})^2 + (\varrho_{11}-\varrho_{33})^2]=0.
\end{eqnarray}
(3)~$ (\varrho_{12}+\varrho_{03})^2  \leq(\varrho_{00}^2+\varrho_{11}^2+\varrho_{22}^2+\varrho_{33}^2)/2- \varrho_{00} \varrho_{22} - \varrho_{11} \varrho_{33}\leq \varrho_{00}^2+\varrho_{11}^2+\varrho_{22}^2+\varrho_{33}^2 - 1/4 $:
\begin{equation}\label{DG_D3}
    D^G(\rho_{AB})=D(\rho_{AB})=2 (\varrho_{12}^2 + \varrho_{03}^2).
\end{equation}
We conclude that $D^G(\rho_{AB})\geq D(\rho_{AB})$ for any $X$ state from Eqs.(\ref{DG_D},\ref{DG_D1},\ref{DG_D2},\ref{DG_D3}).

\section{SUMMARY}\label{s4}
In summary, we have proved GGQD and TQC are the same. Then we obtained analytical formulas of GGQD and GD for two-qubit $X$ states. In addition, we have further found that GD is the tight lower bound of GGQD, which means that GD is a good approximation at least for $X$ states. There are still some interesting opening problems needed to be studied in this respect, such as, are there any analytical expressions of GGQD for qubit-qudit system? Can GD be a tight lower bound of GGQD for any bipartite system? We shall report our research results on these issues later.


\begin{thebibliography}{99}
\bibitem{Nielsen} M. A. Nielsen and I. L. Chuang, Quantum Computation
and Quantum Information (Cambridge University Press, Cambridge,
England, 2000).
\bibitem{pra_40_4277_1989} R. F. Werner, Phys. Rev. A \textbf{40}, 4277 (1989).
\bibitem{rmp_80_517_2008} L. Amico, R. Fazio, A. Osterloh, and V. Vedral,  Rev. Mod. Phys. \textbf{80}, 517 (2008).
\bibitem{rmp_81_865_2009} R. Horodecki, P. Horodecki, M.
 Horodecki, and K. Horodecki, Rev. Mod. Phys. \textbf{81}, 865 (2009).
\bibitem {PRA_72_042316_2005} A. Datta, S. T. Flammia, and C. M. Caves, Phys. Rev. A \textbf{72}, 042316 (2005);
\bibitem{PRL_100_050502_2008} A. Datta, A. Shaji, and C. M. Caves, Phys. Rev.
Lett. \textbf{100}, 050502 (2008).
\bibitem {PRL_101_200501_2008} B. P. Lanyon, M. Barbieri, M. P. Almeida, and A. G. White,
 Phys. Rev. Lett. \textbf{101}, 200501 (2008).
\bibitem {PRL_88_017901_2001} H. Ollivier and W. H. Zurek, Phys. Rev. Lett. \textbf{88},
017901 (2002).
\bibitem{JPA_34_6899_2001} L. Henderson and V. Vedral, J. Phys. A \textbf{34}, 6899
(2001).
\bibitem{PRA_67_012320_2003} W. H. Zurek, Phys. Rev. A \textbf{67}, 012320 (2003).
\bibitem {prl_100_090502_2008} M. Piani, P. Horodecki, and R. Horodecki,  Phys. Rev. Lett.
\textbf{100}, 090502 (2008).
\bibitem {pra_77_022301_2008} S. Luo, Phys. Rev. A \textbf{77}, 022301 (2008).
\bibitem {pra_78_024303_2008} N. Li and S. Luo, Phys. Rev. A \textbf{78}, 024303 (2008).
\bibitem {pra_81_042105_2010} M. Ali, A. R. P. Rau, and G. Alber, Phys. Rev. A \textbf{81}, 042105 (2010).
\bibitem {pra_82_069902_2010}M. Ali, A. R. P. Rau, and G. Alber, Phys. Rev. A \textbf{82}, 069902 (2010).
\bibitem {prl_89_180402_2002} J. Oppenheim, M. Horodecki, P. Horodecki, and R. Horodecki,
Phys. Rev. Lett. \textbf{89}, 180402 (2002).
\bibitem {prl_101_070502_2008} D. Kaszlikowski, A. Sen(De), U. Sen, V. Vedral, and A.Winter,
Phys. Rev. Lett. \textbf{101}, 070502 (2008).
\bibitem {pra_76_032327_2007} N. Li and S. Luo, Phys. Rev. A \textbf{76}, 032327 (2007);
\bibitem {pra_77_42303_2008} S. Luo, Phys. Rev. A \textbf{77}, 042303 (2008).
\bibitem {prb_78_224413_2008} R. Dillenschneider, Phys. Rev. B \textbf{78}, 224413 (2008).
\bibitem {pra_80_022108_2009} M. S. Sarandy, Phys. Rev. A \textbf{80}, 022108 (2009).
\bibitem {prl_105_150501_2010} M. D. Lang and C. M. Caves, Phys. Rev. Lett. \textbf{105}, 150501 (2010).
\bibitem {njp_16_033027_2014} Y. Huang, New J. Phys. \textbf{16}, 033027 (2014).
\bibitem {pra_81_052107_2010} F. F. Fanchini, T. Werlang, C. A. Brasil, L. G. E. Arruda,
and A. O. Caldeira, Phys. Rev. A \textbf{81}, 052107 (2010).

\bibitem {prl_105_095702_2010} T. Werlang, C. Trippe, G. A. P. Ribeiro,
and G. Rigolin, Phys. Rev. Lett. \textbf{105}, 095702 (2010).
\bibitem {pra_82_042316_2010} L. Ciliberti, R. Rossignoli, and N. Canosa, Phys. Rev. A \textbf{82}, 042316 (2010).
\bibitem {pra_81_022116_2010} J. Maziero, T. Werlang, F. F. Fanchini, L. C. C\'{e}leri, and
R. M. Serra, Phys. Rev. A \textbf{81}, 022116 (2010).

\bibitem {pra_83_012327_2011} X.-M. Lu, J. Ma, Z. Xi, and X. Wang, Phys. Rev. A \textbf{83}, 012327 (2011).
\bibitem {pra_84_042313_2011} Q. Chen, C. Zhang, S. Yu, X. X. Yi, and C. H. Oh, Phys. Rev. A
\textbf{84}, 042313 (2011).
%
\bibitem{JPA_45_095303_2012} S. Vinjanampathy and A R P Rau, J. Phys. A: Math. Theor. \textbf{45}, 095303 (2012).
\bibitem {pra_88_014302_2013} Y. Huang, Phys. Rev. A \textbf{88}, 014302 (2013).


\bibitem{prl_105_190502_2010} B. Daki\'{c}, V. Vedral, and \v{C}. Brukner, Phys. Rev. Lett. \textbf{105},190502 (2010).
\bibitem{pra_82_034302_2010} S. Luo and S. Fu, Phys. Rev. A \textbf{82}, 034302 (2010).
\bibitem {pra_85_024102_2012} S. Rana and P. Parashar, Phys. Rev. A \textbf{85}, 024102 (2012).
\bibitem{pra_85_024302_2012} A. S. M. Hassan, B. Lari, and P. S. Joag, Phys. Rev. A \textbf{85}, 024302 (2012).
\bibitem {IJMPB_27_1345020_2012} D. Girolami, R. Vasile and G. Adesso, Int. J. Mod. Phys. B
\textbf{27}, 1345020 (2012).
\bibitem{prl_108_150403_2012} D. Girolami and G. Adesso, Phys. Rev. Lett. \textbf{108}, 150403 (2012).
\bibitem{pra_86_052326_2012} T. Tufarelli, D. Girolami, R. Vasile, S. Bose, and G. Adesso,
 Phys. Rev. A \textbf{86} 052326 (2012).


\bibitem{PRA_84_042109_2011} C. C. Rulli and M. S. Sarandy, Phys. Rev. A \textbf{84}, 042109 (2011).
\bibitem{PRA_87_062339_2013} D. P. Chi, J. S. Kim, and K. Lee, Phys. Rev. A \textbf{87}, 062339 (2013).
\bibitem{pla_377_238_2013} J. Xu, Phys. Lett. A \textbf{377}, 238 (2013).
\bibitem{jpa_45_405304_2012} J. Xu, J. Phys. A: Math. Theor. \textbf{45}, 405304 (2012).
\bibitem{PRA_86_042123_2012} A. Miranowicz, P. Horodecki, R. W. Chhajlany, J. Tuziemski,
 and J. Sperling, Phys. Rev. A \textbf{86}, 042123 (2012).
\bibitem{cpb_22_040303_2013} F. J. Jiang, H. J. L\"{u}, X. H. Yan, and M. J. Shi, Chin. Phys. B  \textbf{22}, 040303 (2013).
\bibitem{jpa_45_345301_2012} A. S. M. Hassan and P. S. Joag,, J. Phys. A: Math. Theor. \textbf{45}, 345301 (2012).
\bibitem{pra_85_022320_2012} F.  Lastra, C. E. L\'{o}pez, L. Roa, and J. C.
Retamal, Phys. Rev. A 85, 022320 (2012).
\end{thebibliography}
\end{document}